\documentstyle[12pt]{article}
\begin{document}
\addtolength{\baselineskip}{.1mm}
\input epsf
\newcommand{\vev}[1]{\langle #1 \rangle}
\def\mapright#1{\!\!\!\smash{
\mathop{\longrightarrow}\limits^{#1}}\!\!\!}
\newcommand{\bigoint}{\displaystyle \oint}
\newlength{\extraspace}
\setlength{\extraspace}{2mm}
\newlength{\extraspaces}
\setlength{\extraspaces}{2.5mm}
\newcounter{dummy}
\newcommand{\be}{\begin{equation}
\addtolength{\abovedisplayskip}{\extraspaces}
\addtolength{\belowdisplayskip}{\extraspaces}
\addtolength{\abovedisplayshortskip}{\extraspace}
\addtolength{\belowdisplayshortskip}{\extraspace}}
\newcommand{\ee}{\end{equation}}
\newcommand{\figuur}[3]{
%\addtocounter{fignum}{1}
%\addcontentsline{lof}{figure}{\protect
%\numberline{\arabic{section}.\arabic{fignum}}{#3}}
%\hspace{-3mm}{\it fig.}\ %\figuurnum.
\begin{figure}[t]\begin{center}
\leavevmode\hbox{\epsfxsize=#2 \epsffile{#1.eps}}\\[3mm]
\parbox{15cm}{\small \bf Fig.\ %\figuurnum:
\it #3}
\end{center} \end{figure}\hspace{-1.5mm}}
\newcommand{\fig}{{\it fig.}\ }
\newcommand{\newsection}[1]{
\vspace{15mm}
\pagebreak[3]
\addtocounter{section}{1}
\setcounter{subsection}{0}
\setcounter{footnote}{0}
\noindent
{\large\bf %\thesection. 
#1}
\nopagebreak
\medskip
\nopagebreak}
\newcommand{\newsubsection}[1]{
\vspace{1cm}
\pagebreak[3]
%\addtocounter{subsection}{1}
%\addcontentsline{toc}{subsection}{\protect
%\numberline{\arabic{section}.\arabic{subsection}}{#1}}
%\noindent{\it \thesection.
%\thesubsection. 
\noindent
{\bf #1}
\nopagebreak
\vspace{2mm}
\nopagebreak}
\newcommand{\ba}{\begin{eqnarray}
\addtolength{\abovedisplayskip}{\extraspaces}
\addtolength{\belowdisplayskip}{\extraspaces}
\addtolength{\abovedisplayshortskip}{\extraspace}
\addtolength{\belowdisplayshortskip}{\extraspace}}
\newcommand{\one}{{\bf 1}}
\newcommand{\zbar}{\overline{z}}
\newcommand{\ea}{\end{eqnarray}}
\newcommand{\is}{& \!\! = \!\! &}
\newcommand{\hf}{{1\over 2}}
\newcommand{\del}{\partial}
%
% 2 by 2 matrices
%
\newcommand{\twomatrix}[4]{{\left(\begin{array}{cc}#1 & #2\\
#3 & #4 \end{array}\right)}}
\newcommand{\twomatrixd}[4]{{\left(\begin{array}{cc}
\displaystyle #1 & \displaystyle #2\\[2mm]
\displaystyle  #3  & \displaystyle #4 \end{array}\right)}}
\newcommand{\low}{{{\rm {}_{IR}}}}
\newcommand{\hi}{{{\rm {}_{UV}}}}
\newcommand{\RR}{{{\rm {}_{R}}}}
\newcommand{\ie}{{\it i.e.\ }}
\newcommand{\gbar}{{\overline{g}}}
\newcommand{\half}{{\textstyle{1\over 2}}}
\newcommand{\tfrac}[2]{{\textstyle{\frac{#1}{#2}}}}
\newcommand{\ttau}{r}
\newcommand{\UU}{U}
\renewcommand{\thesubsection}{\arabic{subsection}}
\begin{titlepage}
\begin{center}

{\hbox to\hsize{
\hfill PUPT-1898}}
{\hbox to\hsize{
\hfill ITFA-99-39}}
{\hbox to\hsize{
\hfill SPIN-1999/29}}

\bigskip

\vspace{6\baselineskip}

{\large \sc On the Holographic Renormalization Group}

\bigskip

\bigskip
\bigskip

{ \sc Jan de Boer${}^{2,3}$, Erik Verlinde$^{1}$, and 
Herman Verlinde${}^{1,4}$}\\[1cm]

{${}^1$ \it Physics Department, Princeton University, Princeton, NJ 08544}\\[3mm]
{${}^2$ \it Spinoza Institute, University of Utrecht,\\  Leuvenlaan 4, 3584, CE Utrecht, The Netherlands}\\[3mm]
%{${}^3$  \it Theory Division, CERN, CH-1211 Geneva 23, Switzerland.}\\[3mm]
{${}^3$   \it Insituut-Lorentz for Theoretical Physics, University of Leiden,
P.O.Box 9506, NL-2300 RA Leiden, The Netherlands.}\\[3mm]
{${}^4$ \it Institute for Theoretical Physics, University of Amsterdam, 
1018 XE Amsterdam}\\[.5cm]

\vspace*{1.5cm}
\large{

{\bf Abstract}\\}

\end{center}
\noindent
We propose a direct correspondence between the classical evolution
equations of 5-d supergravity and the renormalization group
(RG) equations of the dual 4-d large $N$ gauge theory. Using standard 
Hamilton-Jacobi theory, we derive first order flow equations for the
classical supergravity action $S$, that take the usual form of the 
Callan-Symanzik equations, including the corrections due to the 
conformal anomaly. This result gives direct support for the 
identification of $S$ with the quantum effective action of the gauge theory.
In addition we find interesting new relations between the
beta-functions and the counterterms that affect the 4-d cosmological
and Newton constant.  

\end{titlepage}

\newpage
\newcommand{\cto}{\mu}
\newcommand{\bebar}{{\widehat{\pi}}}

\newsubsection{Introduction}

A central element in the correspondence
 between classical 5-dimensional supergravity and
4-dimensional large $N$ gauge theory \cite{juan} is the interpretation of the
extra `radial' 5-th coordinate $\ttau$ with a renormalization group
parameter of the 4-d field theory \cite{gkp} \cite{witten}
\cite{holo}. The radial evolution of the 5-d fields indeed shares many
features with an RG flow 
\cite{extr1} \cite{extr2} \cite{nicketal} 
\cite{porratietal} 
\cite{rg1} \cite{rg2} \cite{rg3} \cite{rg4}, but a complete understanding of
this correspondence still seems to be lacking. The purpose of this 
paper is to fill some of the gaps in this dictionary and to further
clarify the relation between gravity and the renormalization group. 
Concretely, we
will show that the supergravity equations of motion can be
reduced to a flow equation for the classical supergravity
action $S$, which furthermore, in the asymptotic limit, can be cast in 
the form of a standard Callan-Symanzik equation. We regard
this result as additional justification for the identification of the 
supergravity action $S$ with the quantum 
effective action of the dual 4-d gauge theory. We furthermore will 
obtain new interesting relations between the various terms in the
gravitational part of the 4-d effective action, in particular the
terms that affect the Newton constant and the cosmological term. These
relations suggest an intimate connection between the
physics of the renormalization group and the 4-d Einstein equations. 
This point is worked out in more detail in a separate paper \cite{two}. 
Our discussion will be quite general and will not be restricted to a 
particular dual pair of supergravity and (conformal) gauge theory. 
Although our approach is new, there is some overlap between our results and 
earlier papers 
\cite{extr1} \cite{extr2} \cite{nicketal} 
\cite{porratietal} \cite{rg1} \cite{rg2} \cite{rg3} \cite{rg4} 
on the holographic RG-flow.

\newsubsection{The holographic correspondence}

To set notation we begin with a quick review of the basic elements of
the AdS/CFT correspondence \cite{juan} \cite{gkp} \cite{witten}. The
central proposal here is that there exists an exact duality between
large $N$ 4-dimensional ${\cal N}=4$ Yang-Mills theory and type IIB
string theory, or for large 't Hooft coupling, IIB supergravity living
on $S_5$ times $AdS_5$. The supergravity theory contains scalars
$\phi^I$ that represent the couplings of the 4-d gauge theory, and the
propagation of these fields in the $\ttau$-direction is believed to
correspond to the 4-dimensional RG-flow. Since these fields are
coupled to gravity, their stress energy will affect the
shape of the background geometry, which as a result will no longer be
given by the pure AdS form. To study this radial evolution it will be
convenient to choose an analogue of the familiar temporal gauge, in
which the 5-d metric takes the form\footnote{To
re-introduce the lapse and shift functions one can make the substition 
$d\ttau\to Nd\ttau$ and $dx^\mu\to dx^\mu+N^\mu d\ttau$}
\be
\label{gauge}
ds^2= d\ttau^2+g_{\mu\nu}(x,\ttau)dx^\mu dx^\nu.
\ee
The metric $g_{\mu\nu}$ and the fields $\phi^I$ are allowed to be dependent
on all five coordinates $x^\mu$ and $\ttau$.   

We choose a set-up in which the 5-d space has Euclidean signature and
has a boundary at a finite value of $\ttau$. We now define $S[\phi,g]$
as the classical action of the supergravity solution with boundary
values $\phi^I(x)$ for the fields, and $g_{\mu\nu}(x)$ for the
metric. To get a unique configuration for a given choice of $\phi^I$
and $g_{\mu\nu}$ we further impose that the resulting five-dimensional
space, even though it is geometrically different from $AdS_5$, still
has the same topology, namely that of a five-ball $B_5$.  By requiring
that the fields $\phi^I$ and the metric remain regular in the interior
of $B_5$, there is in principle one unique classical solution for
given boundary values of the fields and the metric \cite{witten}.
According to the prescription of \cite{gkp} \cite{witten}, the
classical action $S[\phi,g]$ evaluated on the corresponding 
classical solution
describes the generating function of gauge
invariant operators ${\cal O}_I$ in the gauge theory
\be 
{\large\langle} \, {\cal O}_{I_1}(x_1)\ldots {\cal O}_{I_n}(x_n) \,
{\large\rangle} = 
{1\over{\sqrt {g}}}\,
{\delta \ \ \over \delta\phi^{I_1}(x_1)}\ldots{1\over{\sqrt {g}}}\,
{\delta\ \  \over \delta\phi^{I_n}(x_n)}\, S\, [\, \phi\, ,\, g\, ]. 
\ee
When one puts the fields $\phi^I$ to zero after doing the variation one
computes the correlators of the unperturbed $N\!=\! 4$ supersymmetric 
Yang-Mills theory. One can also put $\phi^I$ equal to a finite value, 
which corresponds to switching on a finite perturbation of the field 
theory. So in principle we could consider all gauge theories that 
can be obtained by a general perturbation of the $N\! = \! 4$ SYM theory.

Keeping the boundary at a finite radial position $\ttau$ amounts to
keeping a finite UV cut-off in the gauge theory. Before one can remove
this cut-off and take the limit $\ttau\to\infty$, one must cancel the
infinities that arise in the classical action of the supergravity
solution. The main source of infinities is the growing metric, or
volume, at infinity, since we have $g_{\mu\nu}\sim e^{2\lambda
r}\hat{g}_{\mu\nu}$, where $\lambda$ denotes the inverse curvature
radius of the asymptotic AdS-space.  One of our goals will be to use
the classical supergravity equations to show that, after subtracting
these infinities, the resulting correlation functions satisfy the
standard Callan-Symanzik equations, identical in form to the one
derived from the gauge theory. The idea behind this correspondence is
that the infinities one finds in taking the AdS-boundary to infinity
can be directly matched those of the quantum field theory calculation.
In both cases the divergences can be canceled by means of a finite
number of counter terms, provided one restricts to renormalizable
perturbations.  On the supergravity side, these renormalizable
perturbations are represented by `tachyonic' scalar fields $\phi^I$
with $4<\lambda^2 m_{{}_I}^2<0$.  These fields decay exponentially as
we approach the boundary at $\ttau=\infty$, and therefore will not
contribute to any additional infinities.  In the following we will
always restrict ourselves to renormalizable and marginal
perturbations.

\smallskip

\newsubsection{Radial evolution}

The 5-d equations of motion in the gauge (\ref{gauge}) can be written
as a Hamilton system, with time replaced by the radial coordinate
$\ttau$. The hamiltonian that generates this radial flow is the analog
of the familiar ADM hamiltonian
\be
H\, =\, \int\! d^4x\, \sqrt{g}\; {\cal H},
\ee
with
\be
{\cal H}
=
\Bigl(\pi^{\mu\nu}\pi_{\mu\nu}\! -{{1\over 3}} \pi^\mu_\mu \pi^\nu_\nu\Bigr)
+ {{1\over 2}} \pi_{I}G^{IJ}(\phi)\pi_{J}
+ {\cal L}(\phi,g).
\ee
Here $\pi_{\mu\nu}$ and $\pi_I$ are the canonical momentum variables conjugate
to $g^{\mu\nu}$ and $\phi^I$, and ${\cal L}$ denotes
the local lagrangian density 
\be
{\cal L}(\phi,g) = V(\phi) + R +
\frac{1}{2} \, \partial^\mu\phi^I G_{IJ}(\phi)\, \partial_\mu \phi^J
\ee
with $V(\phi)$ and $G_{IJ}(\phi)$ the 5-d scalar potential and metric.
Here we have chosen to work in
the 5-dimensional Einstein frame, so that we can use 5-d Planck units
with $\kappa_5=1$.

The 5-dimensional supergravity
equations of motion are implied by the standard Hamilton equations,
%$$\dot{\phi}^I  = {1\over{\sqrt {g}}}\, 
%{\delta \, H \over \delta \pi_{I}},  
%\qquad \qquad\quad \dot{\pi}_I = -{1\over{\sqrt {g}}}\,
%{\delta \, H \over \delta \phi^{I}},
%$$
%for the scalars, and
%$$\dot{g}^{\mu\nu} = {1\over{\textstyle{\sqrt {g}}}}\,
%{\delta \, H \over \delta \pi_{\mu\nu}},\qquad \qquad\quad 
%\dot{\pi}_{\mu\nu}=- {1\over{\textstyle{\sqrt {g}}}}\,
%{\delta \, H \over \delta g^{\mu\nu}},
%$$
%\smallskip  
%\noindent
%for the metric. These equations are to be
supplemented with the additional constraints 
\be
\label{conv}
\nabla^\mu\pi_{\mu\nu}+\pi_I\nabla_\nu\phi^I=0,
\ee
as well as the Hamilton
constraint
\be
\label{His0}
{\cal H}=0,
\ee
which ensures the invariance under local shifts 
$\ttau \to \ttau + \delta \ttau(x)$ of the `equal time slices' $\ttau =$ const.

\newsubsection{Hamilton-Jacobi equation}

Let us now bring the action $S[\phi,g]$ back in to our discussion.
It is a standard fact, well known in classical mechanics, that the value of the
canonical momentum conjugate to $\phi^I$ at a given `time' $\ttau$ is
equal to the functional $\phi^I$ derivative of the classical action $S$,
for the classical solution with boundary value $\phi^I(\ttau)=\phi^I$. 
This fact also holds when the `time'
direction is replaced by a space-like coordinate.  
Calling this momentum variable $\pi_I$, we thus 
have
%\footnote{From now on we will simply denote the
%effective action by $S$, dropping the subscript $ef \! f$.}
\be
\label{lambd}
\pi_{I}={1\over {\sqrt{g}}}{\delta S\over \delta \phi^I}. 
\ee 
This relation is true independent of the initial conditions or
other details of the classical trajectory. Furthermore, the momentum 
$\pi_I$ is related to the flow velocity $\dot{\phi}^I$  by
\be
\label{pil}
%\qquad \qquad \qquad \qquad
\dot{\phi}^I=G^{IJ}\pi_J \, 
%\beta^I \, = \, \dot{\phi}^I.
\ee
where $G_{IJ}$ is again the 5-d scalar metric on the space of couplings. 
Here the dot represents differentiation with respect to the radial coordinate
$\ttau$, even though it is actually a space-like coordinate.

We can do the same for the metric. In classical 5-d gravity, 
the momentum variable
$\pi_{\mu\nu}$ conjugate to $g_{\mu\nu}$ is expressed as 
\be
\label{bmunu}
\pi_{\mu\nu}
=\frac {1}{\sqrt{g}}\frac{\delta S}{\delta g^{\mu\nu}}. 
\ee 
and is related to the flow velocity of $g_{\mu\nu}$ 
via\footnote{With non-trivial lapse and shift $\dot{g}_{\mu\nu}\to
N^{-1}(\dot{g}_{\mu\nu}-\nabla_\mu N_\nu-\nabla_\nu N_\mu)$.} 
\be
\label{pig}
\dot{g}_{\mu\nu}= 2\pi_{\mu\nu}- {2\over 3}
\pi^\lambda_\lambda \, g_{\mu\nu}.
\ee
Combining these four equations gives the first order form of the flow of the
couplings $\phi^I$ and the metric. Given the functional form of the 
classical action $S[\phi,g]$ one can unambiguously compute 
the radial derivative of the couplings and metric in 
terms of their values at that point. 

Let us insert the relations 
(\ref{bmunu}) and (\ref{pig}) into the constraints (\ref{conv}) and 
(\ref{His0}). Equation (\ref{conv}) gives
\be
\nabla^\mu{\delta S\over\delta g^{\mu\nu}}+\nabla_\nu\phi^I{\delta S\over\delta\phi^I}=0.
\ee
This constraint is easily satisfied: it simply 
means that the effective action $S$ is invariant under 4-d coordinate 
transformations. Equation (\ref{His0}) takes the form
\be
\label{ham}
\frac {1}{\sqrt{g}}\left({{1\over 3}}
\Bigl(g^{\mu\nu}\frac{\delta S}{\delta g^{\mu\nu}}\Bigr)^2-
\frac{\delta S}{\delta g^{\mu\nu}}
\frac{\delta S}{\delta g_{\mu\nu}}
-\frac {1}{2} G^{IJ}{\delta S\over \delta \phi^I}
{\delta S\over \delta \phi^J}\right)
= \sqrt{g} {\cal L}(\phi,g).
\ee
This Hamilton-Jacobi constraint will play a central role in the remainder.  
It is important to realize that it is not a 
constraint on the {\it value} of the variations of the action $S$ nor an 
equation of motion of the 4-dimensional fields: instead one must read it 
as a functional differential equation\footnote{It is instructive 
to compare it to the Hamilton-Jacobi equation for a relativistic point 
particle: $({\partial_t S})^2-({\partial_x S})^2=m^2$. We see that scale 
variation plays the role of time differentiation, which presumably is part 
of the reason for why the conformal factor has the `wrong' sign 
in the Einstein action.}
 that determines the 
functional form of the classical action $S$.

\newsubsection{Derivative expansion}.

At the scale of the cut-off $\mu_c\sim e^{\lambda r}$ the action 
$S[\phi,g]$ is non-local. But at an energy scale $\mu<\!<\mu_c$ a 
part of $S$ can be represented as a local action.
Hence, the action $S[\phi,g]$ can in that case be 
be decomposed in a local and a non-local part
\be
\label{deco}
S[\, \phi\, ,\, g\, ] \, = \, S_{\rm loc}\, [\, \phi\, ,\, g\,] 
\; + \; \Gamma\, [\, \phi\, ,\, g\,] 
\ee
with 
\be
S_{\rm loc}\, [\, \phi\, ,\, g\, ] \, =  \;
\int \!\! \sqrt{g}  \, \Bigl(\UU(\phi)  +   
\Phi(\phi) R \, +  \frac{1}{2}\partial^\mu\phi^I M_{IJ}(\phi)\, 
\partial_\mu \phi^J \Bigr),
\ee
where $\UU$, $\Phi$ and $M_{IJ}$  are local functions of the couplings, and
$\Gamma[\phi,g]$ contains all higher derivative and non-local 
terms. Eventually we are interested in the limit $\mu_c\to \infty$, or more precisely, $\epsilon=\mu/\mu_c\to 0$. In that limit the local potential term will
contain quartic divergences, while the other local terms are 
quadratically divergent. In fact, even $\Gamma$ contains divergences of a 
logarithmic type. 
We further note that the local action $S_{\rm loc}$ is
similar in structure as the lagrangian term ${\cal L}$ in the Hamilton 
constraint (\ref{ham}). In fact, as we will see, the different
terms in $S_{\rm loc}$ have a direct relation with the corresponding 
terms in ${\cal L}$. 

The idea of the following calculations will be to insert the expansion
(\ref{deco}) into the Hamilton constraint (\ref{ham}), combine the
contributions on the left hand side that have the same functional form 
as the terms on the right hand side, and require them to cancel. This 
procedure becomes a systematic expansion in the situation where $\mu<\!<\mu_c$,
since the different local and non-local terms can then be 
distinguished by their scaling behavior, up to a redefinition of
$\Gamma\, [\, \phi\, ,\, g\,]$ by finite local terms.
We first will focus on the potential term, which in the limit 
$\mu_c\to \infty$ would be quartically divergent.
The other terms involving $\Phi$ and $M_{IJ}$ will be considered 
in a later section.

By comparing the potential terms on both side, we derive that 
$\UU$ and the 5-d potential term $V$ are related via
\be
\label{fst}
V =  \frac{1}{3}
\UU^2 - \frac{1}{2}\del_I \UU \, G^{IJ} \del_J \UU. 
\ee
Hence $V$ is expressed in terms of $\UU$ in the same way as if
$\UU$ were the superpotential $W$ of the 5-d supergravity.
For supersymmetric flows $\UU$ is indeed
equal to the superpotential.
However, in deriving this expression we have not made use of supersymmetry: 
it is simply a consequence of the standard Hamilton-Jacobi theory 
applied to 5-d gravity, cf. 
\cite{rg4}.  In fact, as we will show in the Appendix, 
different potentials $U$ could lead to the same potential $V$, and 
hence $U$ may in general be different from the superpotential $W$. We 
finally note that
the relation (\ref{fst}) between the 4-d potential $U$ and the 
5d potential $V$ was also found to be a sufficient, and possibly 
necessary condition for the stability of domain wall structures 
in 5-d AdS gravity \cite{rg3} \cite{townsend}.

\newsubsection{Holographic RG-flow}

Let us return to the flow equations (\ref{pil}) and (\ref{pig}).
The field $\phi^I$ and also the metric $g_{\mu\nu}$ play a dual role in
our discussion. On the one had they represent the couplings and the
geometrical background of the gauge theory, while on the other hand they
describe the sources for the local operators ${\cal O}_I$ 
and the stress-energy tensor $T_{\mu\nu}$. In standard quantum field theory
one computes the beta-functions for the coupling constants in a flat background
without any sources. Indeed, to compare our flow equations with
the standard RG-flow we have to consider the theory at a length scale which is
much longer than the cut-off scale, so that at the cut-off scale
the fields practically are independent of the 4-d coordinates. In this limit
the potential term $U$ will dominate over all other terms.

So let us write the flow equation for the scalar fields and the metric 
while keeping only the potential term,
\be
\quad \dot{\phi}^I= G^{IJ}(\phi)\partial_J\UU(\phi)
\qquad \qquad \dot{g}_{\mu\nu} = 
{\frac{1}{3}\, \UU(\phi)} \,g_{\mu\nu}.
\ee
We see that the metric simply rescales. Hence we can solve its
flow equation by the Ansatz
\be
g_{\mu\nu}\, =\, a^2\, \widehat{g}_{\mu\nu},
\ee
where $\widehat{g}_{\mu\nu}$ is independent of $\ttau$ and the prefactor $a$ 
satisfies 
\be
\label{adot}
\dot{a}\, =\, \frac{1}{6}\, U(\phi)\,a.
\ee
Since the parameter $a$ in fact determines 
the physical scale, we now replace  the 
$\ttau$ derivatives in the flow equations
by derivatives with respect to $a$, by using 
the relation (\ref{adot}). In this way we obtain
\be
a{d\over da}{\phi}^I \, =\, \beta^I(\phi)
\ee
where the beta-functions are defined by
\be
\beta^I(\phi)={6\over \UU (\phi)}G^{IJ}(\phi)\partial_J\UU (\phi).
\ee
Thus we see that the holographic RG-flow is in general derived from
a potential.

\newsubsection{Relations for the local terms}

The relation (\ref{fst}) for the potentials $V$ and $U$ can now be 
re-expressed as a condition on the squares of the beta-functions:
\be
{1\over 24}\beta^I G_{IJ}\beta^J=1-{3V\over U^2}.
\ee
Since the left hand side is positive definite, the potential $U$ obeys the
inequality $U^2\geq 3V$, where the equal sign only holds at fixed points of the
RG-flow. The relations between the other local terms in the action are 
also conveniently expressed in terms of the beta-functions.
Comparing the curvature and kinetic terms on the right- and left-hand side 
gives
\be
\beta^K\!\partial_K\Phi = 2\Phi+{6\over \UU},
\ee
and
\be
\beta^K\partial_K M_{IJ}\!-\!\beta^K \partial_I M_{KJ}\!-\! 
\beta^K \partial_J M_{IK}
= 2M_{IJ}+{6\over \UU}G_{IJ}.
\ee
In addition, there is also a term on the left-hand side of (\ref{ham})
proportional to $\nabla^2\phi$ that is not present on the right-hand
side. Requiring the coefficient in front of it cancels gives another
interesting expression for the beta-function in terms of purely 
4-d quantities
\be 
\beta^I=M^{IJ}\partial_J\Phi.  
\ee
In most studies of the AdS/CFT correspondence, the local terms of the
boundary effective action are considered to be non-universal, since they
diverge as we take the limit $\ttau\to\infty$. Here we see, however, that
they contain very important information about the flow of the couplings 
constants, and even conspire in an interesting way to make up the lagrangian
term ${\cal L}$ of the 5-d gravity theory.

\newsubsection{Callan-Symanzik equation}

As was mentioned before, 
the action $S$ represents the generating function of correlation functions 
of local operators ${\cal O}_I$ in the gauge theory. In fact, 
the information about 
$n$-point functions of operators ${\cal O}_I$ at different points 
is all contained in the non-local part of the action $\Gamma$.
We already established various relations between the
local terms. Following the same strategy we will now derive a flow
equation for the correlation functions that exactly take the form
of Callan-Symanzik equations. 
 
The computation proceeds as follows: we again insert the decomposition 
(\ref{deco}) into the constraint (\ref{ham}), and drop all local terms
with two or fewer derivatives.
At the order we are interested in, there are cross terms that involve
(functional derivatives of) the potential $\UU$ and the non-local
effective action $\Gamma$. At that same order one also finds 
curvature squared terms and products of the curvature
with space-time derivatives of the scalar fields. Here we will not
write these terms explicitly.  The relation we then find is
\be
\label{loflo}
{1\over \sqrt{g}}
\Bigl(g^{\mu\nu}{\delta\ \over \delta g^{\mu\nu}}-
\beta^I(\phi){\delta \ \over\delta \phi^I}\Bigr)\, \Gamma\, [ \, \phi, \, g \, ]
\; =\ \mbox{4-derivative
terms} 
\ee
In order to derive the Callan-Symanzik equations for expectation
values of local operators we vary this relation with
respect to fields $\phi^{I}$. After doing the variations, the 
fields are put to their constant average value given by 
the couplings of the gauge theory. We further take the metric
to be of the form $g_{\mu\nu}=a^2\eta_{\mu\nu}$, where $a$ is
$x^\mu$-independent. The 4-derivative terms will drop out 
after this step, and play no role as long as one considers operators
${\cal O}_I$ at different points in space. Finally, we integrate
the resulting expression over all of space and replace the functional
derivatives by ordinary derivatives by using the definitions
\be
\int%\!d^4x\,
g^{\mu\nu}\!{\delta\ \over \delta g^{\mu\nu}}=a{\partial\ \over\partial a},
\qquad\qquad\int {\delta \ \over\delta \phi^I}={\partial\ \over \partial\phi^I}
\ee
In this way one derives after some straightforward algebra from (\ref{loflo})  
the standard form of the Callan-Symanzik equations
\be
\label{CalSym}
\Bigl(a{\partial\ \over \partial a}-\beta^I \partial_I\Bigr){\large\langle}
{\cal O}_{I_1}(x_1)\ldots {\cal O}_{I_n}(x_n){\large\rangle} -
\sum_{i=1}^n\gamma\mbox{\raisebox{-2pt}{${}_{I_i}{}^{\!\! J_i}$}}\,
\langle{\cal O}_{I_1}(x_1)..{\cal O}_{J_i}(x_i).. {\cal O}_{I_n}(x_n)\rangle
\, = \, 0
\ee
where\footnote{ By an appropriate choice of contact term, the ordinary  
$\phi^I$ derivative is turned in a covariant derivative $\nabla_I$ that is 
defined in terms of the metric $G_{IJ}$. This ensure that the whole formalism
stays covariant under field redefinitions.} 
\be
\gamma\mbox{\raisebox{-2pt}{${}_I{}^J$}}=\nabla_I\beta^J 
\ee
represent the anomalous scaling 
dimensions of the operators ${\cal O}_I$.
We should note, however, that this equation
is derived still with a finite cut-off. We will describe in a moment how
one can remove the cut-off and obtain the C-S equation for the full 
renormalized $n$-point functions.

\newsubsection{The conformal anomaly}

In order to recover the conformal anomaly, let us relax the condition 
that the metric is flat, and use the constraint
(\ref{ham}) to compute an expression for the trace of the finite part of the 
expectation value of the stress tensor:
\be
\langle T_{\mu\nu}\rangle= 
{1\over \sqrt{-g}}{\delta\Gamma\over \delta g^{\mu\nu}}
\ee
The calculation is again similar as outlined above.
We find that the trace anomaly takes form
\be
\label{cs}
\langle \, T \, \rangle \, = \, \beta^I \partial_I \Gamma \, +\, c \,
R^{\mu\nu} R_{\mu\nu} - d\, R^2.  
\ee 
where $T = T^\mu_\mu$, and the curvature squared terms simply arise due 
to the square of the Einstein term in the action $S_{\rm loc}$. 
We recognize this equation as a standard type
expression as dictated by the broken scale invariance of the effective 
action $\Gamma$ and the trace anomaly relation.\footnote{For a general 4-d 
conformal field theory one would also have a term given by the square of 
the Weyl tensor. In order to reproduce such a term one would have 
to add it by hand to the Hamilton constraint, as a higher order 
correction to the 5-d Einstein action.} 
The coefficients $c$ and $
d$ are given in terms of $\Omega$ and $\Phi$ by
\be 
c \, =\, {6\Phi^2 \over \UU}, \qquad \quad d = {2\over\UU}
\Bigl( \Phi^2 \!- {3\over 2}\partial_I \Phi\, G^{IJ}\!  \partial_J \Phi\Bigr).
\ee 
We can perform a quantitative check on these coefficients, by going to
the fixed point situation, where all $\del_I$ derivative vanish. A
simple calculation shows we can then express $c$ and $d$ in terms of
the 5-d potential term $V$ as $c = 3d =
(V/ 3)^{-3/2}={\lambda^3/ 8}$, 
which reproduces the expression of the holographic Weyl
anomaly obtained in \cite{kostas}. 
The function $c$ is the analogue of the central charge of the 4-d QFT,
and has been proposed as a candidate $C$-function in
\cite{porratietal} and \cite{nicketal}.

\newsubsection{Removing the UV cut-off}

So far we studied the RG-flow of the couplings, action, and
correlation functions as a function of the UV cut-off.
To make contact with more standard type RG-equations, we have to 
follow the usual renormalization procedure of cancelling divergences,
introducing renormalized couplings, and then sending the cut-off to infinity. 
Here we will outline how this procedure works in the present context.
In particular we will consider how (\ref{CalSym})
can be rewritten as the standard Callan-Symanzik equation for the renormalized
$n$-point function.

Removal of the cut-off amounts to taking the limit 
$\ttau\to\infty$. A practical way to describe this limit is to write the 
metric and the couplings as 
\be
\label{gR}
g_{\mu\nu}=\epsilon^{-2}g^{\RR}_{\mu\nu}
\ee
and
\be
\label{phiR}
\phi^I=\phi^I(\phi_\RR,\epsilon)
\ee
where $g_{\mu\nu}^{{R}}$ 
and $\phi^I_\RR$ are the renormalized metric and 
couplings which are kept fixed as we send $\epsilon\to 0$.  
The relation between the bare
couplings $\phi^I$ and the renormalized couplings $\phi^I_\RR$ is obtained by 
integrating the RG-flow 
\be
\label{flowup}
\qquad \qquad \epsilon \,
{\partial \phi^I \over \partial\epsilon} %(\phi_\RR,\epsilon)
\, =  \, \beta^I(\phi) % \Bigl(\phi^I(\phi_\RR,\epsilon)\Bigr),
\qquad\qquad 
\phi^I=\phi^I_\RR \qquad \quad {\mbox{at}}\ \epsilon \! = \! 1
\ee
Form the supergravity perspective, this procedure for
introducing the renormalized couplings means the following. 
Consider the unique classical supergravity trajectory with asymptotic 
boundary conditions specified by the `bare' fields $(\phi^I,g)$. The 
renormalized fields $(\phi^I_\RR, g_\RR)$ then represent
the values of the scalar field on this trajectory
at some finite value of the scale factor, corresponding to some
fixed RG scale.

As we consider only relevant perturbations, the couplings $\phi^I$ will 
actually go to zero as $\epsilon^{\lambda_I/\lambda}$, as we take the 
limit $\epsilon\to 0$ while keeping $\phi_\RR$ fixed. 
The metric $g_{\mu\nu}$ on the other hand diverges. Hence 
we still find, upon
inserting the expressions (\ref{gR}) and (\ref{phiR}) into the action 
$S$, that the various terms contained in $S_{\rm loc}$  diverge in 
the limit $\epsilon\to 0$. For example, the potential term $U$ will
in general be quartically divergent, while $\Phi$ and $M_{IJ}$ contain
at most quadratic divergences. Even the term $\Gamma$ has a potential
logarithmic divergence that has to be removed in the renormalization procedure.
This can be done by adding appropriate counterterms. The
renormalized effective action $\Gamma_\RR$ is defined by
\be
\Gamma_\RR[\phi_\RR,g_\RR]=\lim_{\epsilon\to 0}
\Gamma\mbox{\raisebox{-2pt}{${}_{\rm \! finite}$}}
[\, \phi(\phi_\RR,\epsilon),\, \epsilon^{-2}g_\RR]
\ee
where $\Gamma_{\rm finite}$ is obtained from $\Gamma$ by 
subtracting its divergent part.

We now would like to show that the action $\Gamma_\RR$ again 
satisfies a similar Callan-Symanzik equation as before, but 
now expressed in terms of the renormalized couplings and metric
\be
\label{loflo2}
{1\over \sqrt{g}}
\Bigl(g^{\mu\nu}_\RR{\delta\ \over \delta g_\RR^{\mu\nu}}-
\beta^I_\RR (\phi^I_\RR )
{\delta \ \over\delta \phi_\RR^I}\Bigr)\, \Gamma_\RR\, [ \, \phi_\RR, \, g_\RR \, ]
\; =\ \mbox{\rm local terms} 
\ee
The derivation of this relation is basically a change of variables from
$\phi^I$ to $\phi^I_\RR$. The beta-functions behave as vector fields on
the space of couplings, and simply transform accordingly. This can be seen
as follows. Since the relation between $\phi^I$ and
$\phi^I_\RR$ can also be obtained by flowing down,
the beta-function $\beta^I_\RR$ should be such that
the variation of $\phi^I$ due to an infinitesimal shift in the
cut-off $\epsilon$ while keeping $\phi^I_\RR$ constant,
can be exactly compensated by an infinitesimal RG transformation
of $\phi^I_\RR$:
\be
\label{flowdown}
\epsilon\, {\partial \phi^I \over\partial\epsilon} \, =\, 
\beta^J_\RR \, 
{\partial\phi^I\over \partial \phi^J_\RR}.
\ee
By comparing this relation with (\ref{flowup}) one indeed finds the usual
transformation properties of a vector field. This implies that we can directly
make the substitution
\be
\beta^I
{\delta\ \over \delta\phi^I}=\beta^I_\RR{\delta\ \over\delta\phi^I_\RR}
\ee 
Furthermore, since the metric $g^{{}^{\rm R}}_{\mu\nu}$ and $g_{\mu\nu}$ are related by a 
constant rescaling, one can simply relate their functional derivatives as well.
Finally, the operators ${\cal O}_I$ also have to be renormalized, in order for
their correlators to stay finite. We have 
\be
{\cal O}^{{}^{\rm R}}_I\, = \, {\cal O}_J \, {\partial\phi^J\over
\partial \phi^I_\RR}
\ee
With these substitutions it is straightforward to recover the 
Callan-Symanzik equations for all renormalized $n$-point functions.
\be
\label{CalSymR}
\Bigl(a{\partial\ \over \partial a}-\beta_\RR^I 
\partial_I\Bigr){\large\langle}
{\cal O}^{{}^{\rm R}}_{I_1}(x_1)\ldots {\cal O}^{{}^{\rm R}}_{I_n}(x_n)
{\large\rangle}
-
\sum_{i=1}^n(\gamma_\RR)\mbox{\raisebox{-2pt}{${}_{I_i}{}^{\!\! J_i}$}}\,
\langle{\cal O}^{{}^{\rm R}}_{I_1}(x_1)..{\cal O}^{{}^{\rm R}}_{J_i}(x_i).. {\cal O}_{I_n}^{{}^{\rm R}}(x_n)\rangle\, = \, 0
\ee

The above procedure for defining the renormalized beta-functions is
quite standard. As presented, however, it may not be entirely obvious
that the beta-functions at a given scale $\mu$ can in fact be
extracted from just the knowledge of the UV physics at smaller
scales. To see how this arises from the supergravity perspective,
recall that $\phi_\RR^I$ represents the value of the scalar field on a
specified classical supergravity trajectory.  Once we know the values
of the bare fields as well as the initial velocities at the cut-off scale,
we can integrate inwards and obtain the renormalized
beta-functions $\beta^I_\RR(\phi_\RR)$ as the radial velocities (as
measured relative to the overall scale of the metric) of the fields at
this point. This procedure requires knowledge of the supergravity
solution in the outside region only.

\newsubsection{Discussion and outlook}

In this paper we have shown that the Hamilton-Jacobi equations for
5d Einstein gravity can be written in the form of first order 
RG-flow equations. These results can in principle be applied to
every 4d gauge theory that can be represented as a relevant
or marginal perturbation of 
a large $N$ superconformal field theory, or any of its variations for which 
an AdS/CFT correspondence has been established. 
Also, generalizations to other dimensions are straightforward.
In deriving the 
Callan-Symanzik equations we made use of the fact that asymptotically the
metric and scalar fields behave as in AdS-space, but the Hamilton-Jacobi
equations themselves are valid in much more general situations.
For example, it is conceivable that the RG-flow of practically any 
large $N$ gauge theory is, in leading order in $1\over N$, described by
classical supergravity equations of this type. 

The relation between the 5d Einstein equation, the RG-flow and 
Callan-Symanzik equation, as presented in the main text, 
was derived in the special gauge with fixed `lapse' and
`shift' functions. Reintroducing the lapse and shifts would
presumably lead to RG-equations that are invariant under 
local redefinitions of the scale as well as scale-dependent 
coordinate transformations. Together these invariances reflect the 
underlying 5-d general coordinate invariance of the dual 
supergravity theory.

It would be interesting to identify the modifications of our equations
that are required to incorporate $1/N$ corrections. The
Hamilton-Jacobi equations can be considered as the classical limit of
the quantum Wheeler-DeWitt equation, which (when the WDW wave-function
is written as $e^{{i\over \hbar}S}$) contains an additional term proportional
to a second order variation of the action $S$. The resulting equation
has a striking similarity with Polchinski's version of the exact
renormalization group \cite{Polch}.  We believe that this is not a
coincidence, since it should describe the RG-flow of the regularized
(through string theory) quantum supergravity theory.

The internal 5-manifold that features in the usual AdS/CFT framework
has not played an explicit role in our discussion, but its 
geometry is implicitly present in the form of the scalar fields $\phi^I$ 
and the metric $G_{IJ}$. So the RG-flow equations in principle also
describe the behavior of this manifold as one flows towards the infra-red.

Finally, it is an interesting question whether the holographic 
description of the RG-flow that we presented in this paper also 
applies to 4-d gauge theories that are coupled to gravity.
It has been argued in \cite{hv} that one needs to
replace the asymptotic AdS geometry be a compact internal manifold,
as would naturally arise in warped compactifications that give rise to  
4-d gauge theories with gravity. This leads to
an interesting interplay between the 4-d Einstein equations and
the RG-flow equations, that in particular seems to shed new light on the role
of the cosmological constant. This direction is further investigated
in \cite{two} . 

\bigskip

\bigskip
\bigskip

{\noindent \sc Acknowledgements}

This work is supported by NSF-grant 98-02484, a Pionier fellowship
of NWO, the Packard foundation and the stichting FOM.
We would like to thank V. Balasubramanian, R. Dijkgraaf,
S. Kachru, I. Klebanov, G. Lifschytz,
V. Periwal, A. Polyakov, L. Randall, C. Schmidhuber, E. Silverstein,
K. Skenderis for helpful discussions.

\bigskip

\renewcommand{\theequation}{A.\arabic{equation}}
\setcounter{equation}{0}

\bigskip

\noindent
{\bf Appendix: Comment on the Hamilton-Jacobi relation for the 
potential.}\\

In this Appendix we study the relation between the potentials
$U$ and $V$ that follows from the Hamilton-Jacobi equation in more detail.
For simplicity we go to a basis in which the metric $G_{IJ}$ is given by 
$\delta_{IJ}$, so that we can write the relation as
\be
\label{A1}
{1\over 3}U^2-{1\over 2}(\partial_I U)^2=V
\ee 
One might wonder whether any potential $V$ can be written in this form.
Our derivation of this relation did not assume any special properties of $V$,
except the existence of a classical solution that can be extended from the
boundary to the interior. This of course puts restrictions on the
potential $V$, and apparently forces it to be of the supersymmetric form.
We will show in this appendix that the 4d potential $U$ is for a large part,
but not entirely, determined by the 5d potential $V$.

Let us expand the 5-d potential in powers of $\phi^I$ as
\be
\label{A2}
V\,= 12\lambda^2-{1\over 2} 
m^2_{I}\phi^I\phi^I+g_{IJK}\phi^I\phi^J\phi^K\ldots
%+h_{IJKL}\phi^I\phi^J\phi^K\phi^L
\ee
where we used the freedom to shift the fields to remove a possible linear term
in $\phi^I$.  To obtain a solution to (\ref{A1}) for the 4-d potential,
we first try a similar expansion.
\be
\label{A3}
U\,= 6\lambda+{1\over 2}\lambda_{I}\phi^I\phi^I+
\lambda_{IJK}\phi^I\phi^J\phi^K
%+d_{IJKL}\phi^I\phi^J\phi^K\phi^L
\ee
Here we already fixed the constant term so that it matches with that 
of $V$. The beta-functions derived from this potential are
\be
\beta^I = (4-\Delta_I) \phi^I + c^I{}_{JK} \phi^J \phi^K
\ee
where
\be
\label{A4}
\Delta_I = \, 4-{\lambda_I\over \lambda} \qquad \qquad \ 
c_{IJK} = \, {3\over \lambda} \, \lambda_{IJK}
\ee 
are the scaling dimensions and operator product coefficients of the 
operators $O_I$ corresponding to the couplings $\phi^I$.
Inserting both expansions (\ref{A2}) and (\ref{A3})
in to (\ref{A1}) then gives the relation
\be
\lambda_I^2+4\lambda \lambda_I=m^2_I
\ee
which upon inserting the relation (\ref{A4}) is recognized as standard
relation between the scale dimensions $\Delta_I$ of the 4-d couplings
and the corresponding masses $m_I$ of the 5-d fields \cite{gkp}
\cite{witten}.  Notice that this relation implies that the 5-d
potential must in fact satisfy the unitarity bound $m^2_I \geq
-4\lambda^2$. If this inequality is violated, there are no
bounded solutions that extend all the way to the asymptotic boundary.

By looking at the next order in the $\phi^I$ expansion we obtain the
relation
\be
(\lambda_I+\lambda_J+\lambda_K-4\lambda) \lambda_{IJK}= g_{IJK},
\ee
which via (\ref{A4}) expresses the operator product coefficients
$c_{IJK}$ in terms of the cubic term in $V$. Note that this expression 
degenerates in the case that $\Delta_I + \Delta_J + \Delta_k = 8$.
The interpretation of this will become clear when we look at deformations
of $U$ that preserve the relation (\ref{A1}) with $V$.

An infinitesimal variation $\delta U$ preserves (\ref{A1}) if it
satisfies the linear relation
\be
4 U \delta U - 6 \partial_I U \partial_I(\delta U) = 0,
\ee 
or
\be
(4- \beta^I \partial_I ) \delta U = 0.
\ee
These equations tell us that those terms in $U$ that have total dimension $4$ 
are not determined by the Hamilton-Jacobi constraint. It is interesting
to note that these are precisely those terms that remain finite in the 
continuum limit. Therefore, it appears that the Hamilton-Jacobi relation
only constrains the divergent terms of the potential $U$, but not the 
finite part.

\bigskip

\bigskip
\renewcommand{\Large}{\large}

\end{document}